\documentclass[twocolumn,prl, aps,superscriptaddress,showpacs]{revtex4-1}

\usepackage{mathptmx}
\usepackage{subfigure}
\usepackage{dcolumn}
\usepackage{amsmath,amssymb}
\usepackage{bm}
\usepackage{color}
\usepackage{latexsym}
\usepackage{epstopdf}
\usepackage{color}
\usepackage[english]{babel}
\usepackage{latexsym}

\usepackage{psfrag,graphicx} 
\usepackage{epsf} 
\usepackage{subfigure} 
\usepackage{amsmath} 
\usepackage{amssymb} 
\usepackage{amsfonts}
\usepackage{bm}
\usepackage{natbib}
\usepackage{epstopdf}
\usepackage{appendix}

\definecolor{mygrey}{gray}{0.35}
\definecolor{myblue}{rgb}{0.2,0.2,0.8}
\definecolor{myzard}{cmyk}{0,0,0.05,0}
\definecolor{mywhite}{rgb}{1,1,1}
\definecolor{myred}{rgb}{1,0.,0.3}

\usepackage[colorlinks=true,citecolor=myblue,linkcolor=myred]{hyperref}

\def\be{\begin{equation}}
\def\ee{\end{equation}}
\def\ba{\begin{align}}
\def\enda{\end{align}}
\def\bi{\begin{itemize}}
\def\ei{\end{itemize}}

 \def\ee{\mathord{\rm e}}

 \def\ee{\mathord{\rm e}}

\renewcommand{\ee}{{\rm e}}

\def\beq{\begin{equation}}
\def\beq{\begin{equation}}
\def\eeq{\end{equation}}
\def \bml{\begin{multline}}
\def \eml{\end{multline}}
\def \bea{\begin{eqnarray}}
\def \eea{\end{eqnarray}}

\newcommand{\bla}[1]{\left(#1\right)}
\newcommand{\blb}[1]{\left[#1\right]}

\begin{document}

\title[Short Title]{Continuous dynamical decoupling utilizing time-dependent detuning}
\author{I. Cohen}
\author{N. Aharon}
\author{A. Retzker}
\affiliation{Racah Institute of Physics, The Hebrew University of Jerusalem, Jerusalem 91904, Givat Ram, Israel,}

\pacs{ 03.67.Ac, 37.10.Vz, 75.10.Pq}

\begin{abstract}
{Resilience to noise and to decoherence processes is an important ingredient for the implementation of quantum information processing, and quantum technologies. To this end, techniques such as pulsed and continuous dynamical decoupling have been proposed to reduce noise effects. In this paper, we suggest a new approach to implementing continuous dynamical decoupling techniques, that uses an extra control parameter; namely, the ability to shape the time dependence of the detuning. This approach reduces the complexity of the experimental setup, such that we are only left with noise originating from the frequency of the driving field, which is much more robust than the amplitude (Rabi frequency) noise. As an example, we show that our technique can be utilized for improved sensing.}
\end{abstract}
                                            
\maketitle

\section{introduction}
Pulsed dynamical decoupling \cite{Hanh1950,Ban1998jmo,Viola1998pra,Lidar_CDD,Viola1999,Uhring,Yang2011,Souza2012phil,Casanova2015pra} has been used to compensate for fundamental noise sources in different quantum architectures. An alternative approach to the pulsed scheme is continuous, where the system is driven with a protecting dressing field for the entire duration of the experiment \cite{Viola2003prl,Lidar2004pra,Fanchini2007pra,Kurizki,Timoney,Winni2013PRL, Winni_gate, Nati, Mkhitaryan2015scire, Bermudez2012PRA, Tan2013PRL}. In continuous dynamical decoupling an energy gap is opened in the dressed state basis, protecting against the first order contribution of slowly varying noise in a perpendicular direction. However, the new energy gap suffers from Rabi frequency fluctuations, resulting in additional noise which is not compensated for. One way to overcome this problem is known as concatenated continuous dynamical decoupling \cite{Cai2012njp,Lemmer2013njp,Itsik2015NJP,Itsik_Haldane1,Itsik_Haldane2, Cai2015prl}, in which a smaller perpendicular energy gap is opened iteratively in each stage to compensate for the Rabi frequency noise of the previous dressing field. As a result we are left with the Rabi frequency noise of the last dressing fields, which is reduced. However, this comes at the high cost of the complexity of operating with many driving fields, and the resulting reduced energy gap, which forces us to operate slower than it; i.e., all the quantum operators must be slower than the last energy gap we opened.  

In this manuscript we present an alternative approach to use continuous dynamical decoupling which eliminates the difficulties associated with the regular concatenated approach: the complexity, and the reduced protecting energy gap. To this end, rather than using many driving fields, we propose using only a single driving field as protection from the slowly varying noise in a perpendicular direction. To compensate for the Rabi frequency noise of this driving field, we suggest utilizing the arbitrary waveform generator (AWG), to add a time dependent phase to the driving field, yielding a time-dependently detuned driving field. This time-dependent detuning behaves like the second driving field of the concatenated dynamical decoupling scheme; but, due to the enhanced accuracy of the AWG, the noise originating from the detuning is assumed to be negligible, thus removing the need to refocus it. 

\section{Regular concatenated dynamical decoupling}
\begin{figure}
   \centering
  \includegraphics[width=0.45\textwidth]{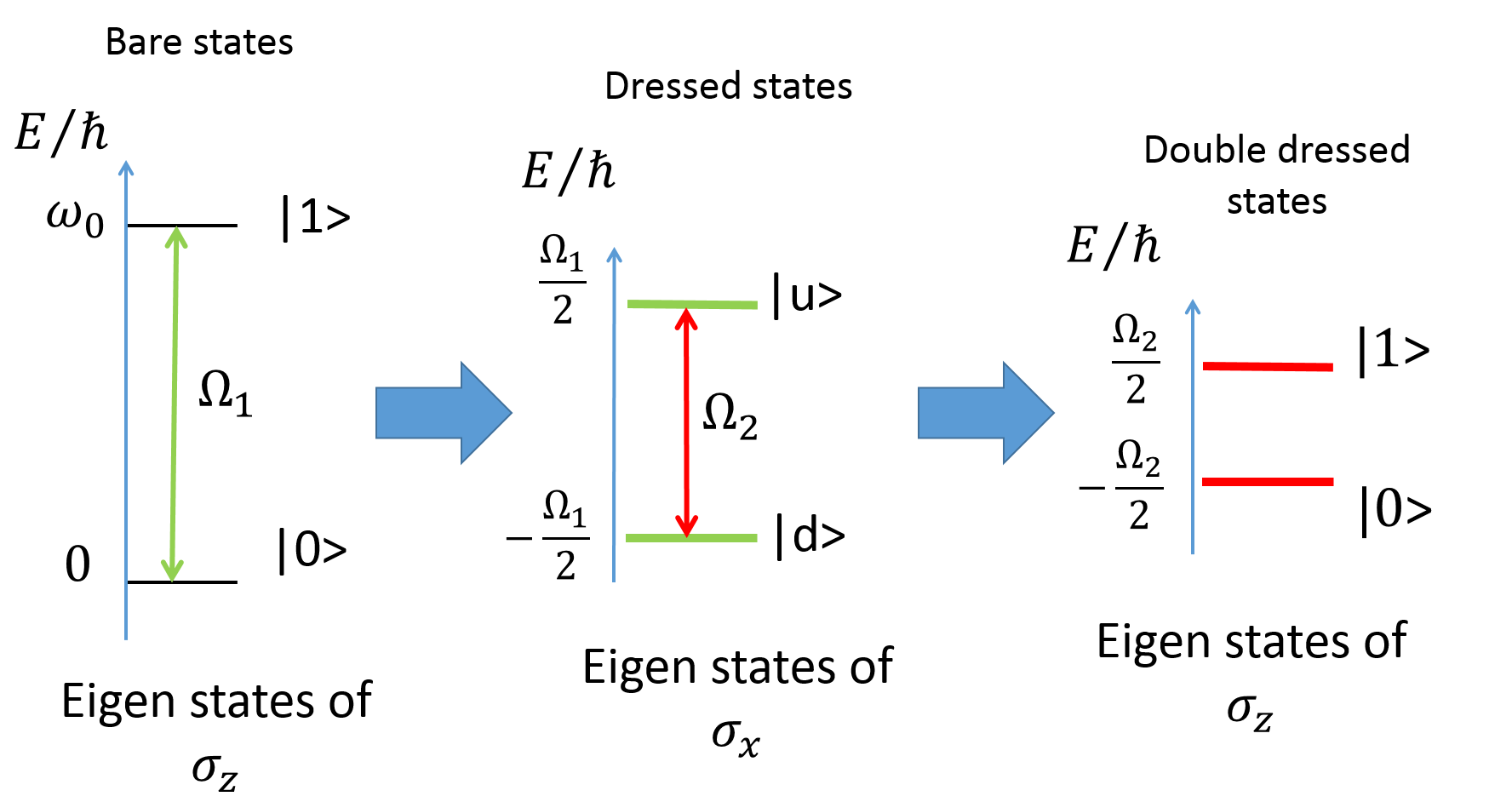}
  \caption{{\bf Regular concatenated dynamical decoupling scheme.} By applying a resonant driving field (Eq. \ref{first_drive}) we open an energy gap in the dressed state basis (Eq. \ref{first_gap}), which is perpendicular to the noise in the bare basis. Thus, the noise in the bare basis is suppressed. To suppress the Rabi frequency (Eq. \ref{second_drive}), a second driving field is used to open the protecting energy gap in the double dressed state basis (Eq. \ref{second_gap}). Thus, we are left with the reduced Rabi frequency noise of the second driving field.
 }
    \label{regular}
\end{figure}

We consider a two-level system with an ambient magnetic field noise in the $z$ direction
\beq
H=\frac{\omega_0}{2} \sigma_z +\delta B(t) \sigma_z.
\eeq
To compensate for this noise, which causes dephasing, we can use the continuous dynamical decoupling approach, in which we drive the system resonantly, in a perpendicular direction to the noise (Fig. \ref{regular}): 
\beq
H_{d1}=\Omega_1 \sigma_x \cos \bla{\omega_0 t}.
\label{first_drive}
\eeq
Therefore, in the rotating frame with the bare energy gap, after using the rotating wave approximation (RWA), $\Omega_1 \ll \omega_0$, the system is described by (Fig. \ref{regular})
\beq
H_{I}=\frac{\Omega_1}{2}\sigma_x +\delta B(t) \sigma_z.
\label{first_gap}
\eeq
In the limit where the DC power spectrum of the ambient magnetic field noise is small compared to the Rabi frequency, $\sqrt{S_{\delta B(t)}(0)} \ll \Omega_1$, %and assuming 1 over $f$ noise, 
thus we are decoupled from this noise to the first order, and we are left with a negligible contribution of the power spectrum of $S_{\delta B(t)}(\Omega_1) \ll S_{\delta B(t)}(0)$. In other words, the ambient magnetic field noise is suppressed by the energy gap that is opened in the dressed state basis. However, the protecting field has its own Rabi frequency noise $\delta \Omega_1(t) \sigma_x$, which causes dephasing as well, thus reducing the coherence time. To compensate for the Rabi frequency noise, Cai {\it et al.} \cite{Cai2012njp} proposed driving the system with a second driving field, in a perpendicular direction to the first one, (Fig. \ref{regular}) e.g.:
\beq
H_{d2}=\Omega_2 \sigma_z \cos \bla{\Omega_1 t}.
\label{second_drive}
\eeq
Therefore, in the two rotating frames, the first one with the bare energy gap, and the second one with the dressed state energy gap, we obtain (Fig. \ref{regular})
\beq
H_{I_2}=\frac{\Omega_2}{2}\sigma_z +\delta \Omega_1(t) \sigma_x,
\label{second_gap}
\eeq
after using the RWA, $\Omega_2 \ll \Omega_1$. Similarly to the above arguments, the new energy gap in the double-dressed state basis compensates for the Rabi frequency noise of the first protecting field, and we are left with the reduced Rabi frequency noise of the second driving field $ \delta \Omega_2(t) \ll \delta \Omega_1(t)$. 

Now the concatenation of these two protecting driving fields can be considered for more driving fields by induction. Each additional driving field will further reduce the dephasing noise. However, the drawbacks of this scheme are the complexity of operating with many driving fields, and the need to carry out additional quantum operations (e.g. quantum gates) slower than the last protecting energy gap. In what follows we describe an improved approach, in which these drawbacks are eliminated.

\section{Continuous dynamical decoupling with a time dependently detuned driving field}
\begin{figure}
   \centering
  \includegraphics[width=0.45\textwidth]{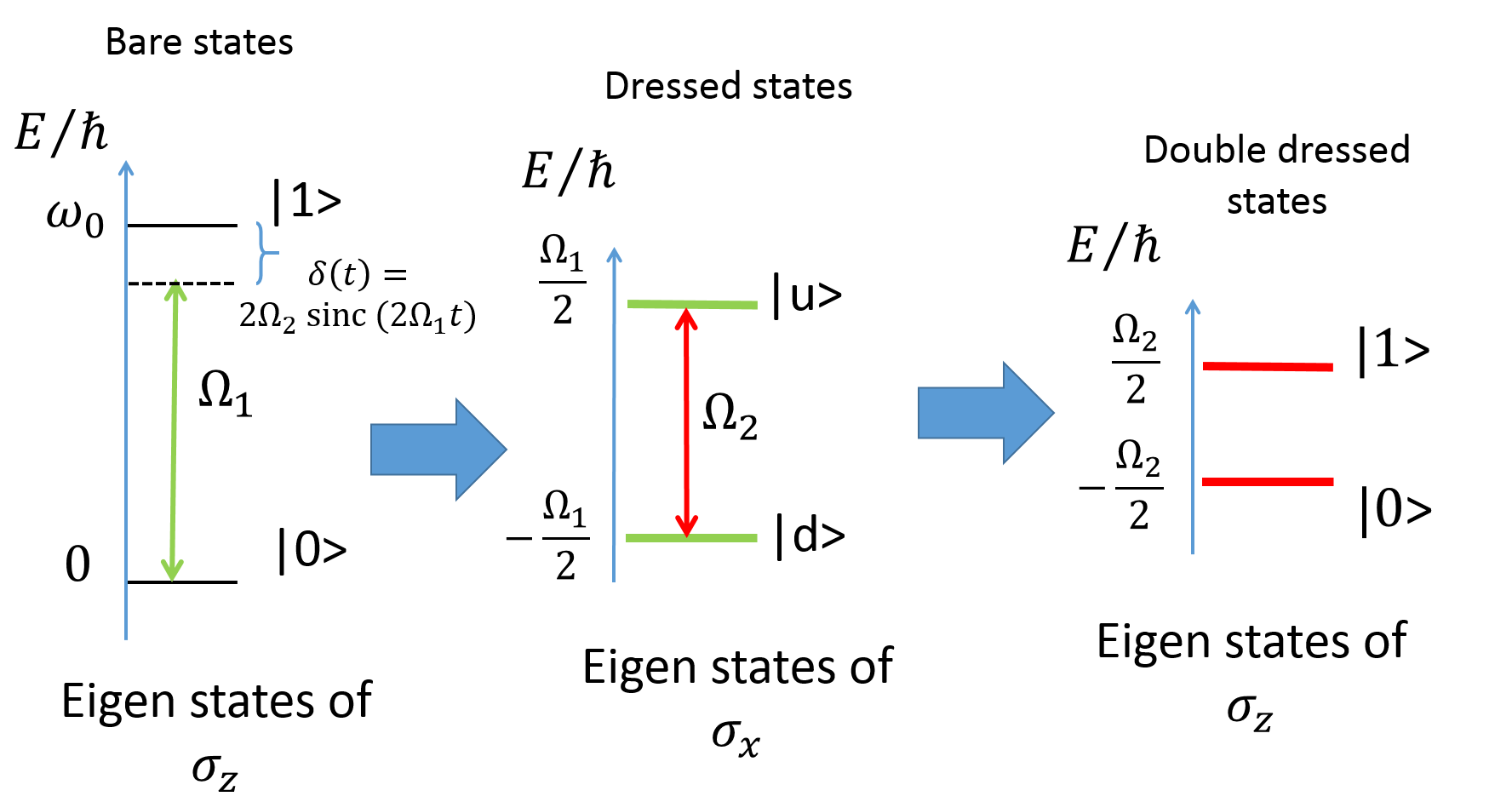}
  \caption{{\bf New continuous dynamical decoupling scheme.} By applying a single, time-dependently detuned driving field (Eq. \ref{time-dependent}) we open both an energy gap in the dressed state basis, and in the double dressed state basis (Eq. \ref{1}), to suppress the noise in the bare basis, and the noise in the dressed state basis, respectively. Now, the fluctuations in the double dressed state basis are assumed to be further reduced compared to the regular scheme (Fig. \ref{regular}), since they originate from the highly accurate time separation of the AWG rather than the driving field amplitude noise.
 }
    \label{time_dependent_concatenation}
\end{figure}

Rather than using many driving fields in order to be left with a reduced dephasing noise, we suggest driving the system with a only single driving field. Utilizing an AWG, a signal having a very accurate time dependent phase $\phi(t)$ can be generated. This results in a time dependently detuned driving field, whose Rabi frequency fluctuates $\Omega_1+\delta \Omega_1(t)$ (Fig. \ref{time_dependent_concatenation}) 
\bea
H=\frac{\omega_0}{2} \sigma_z +\delta B(t) \sigma_z 
+ \blb{\Omega_1+\delta \Omega_1(t) }\sigma_x \cos \bla{\omega_0 t+\phi(t)},
\label{time-dependent}
\eea
with
\beq
\phi(t)=2\frac{\Omega_2}{\Omega_1}\sin\Omega_1 t
\eeq
% $2\Omega_2\mbox{sinc}\Omega_1 t$ 
As will be shown below, the Rabi frequency of the driving field $\Omega_1$ suppresses the ambient magnetic field noise $\delta B(t)\sigma_z$, whereas the time dependent phase $\phi(t)$ suppresses the Rabi frequency noise $\delta \Omega_1(t)\sigma_x$. 

By moving to the interaction picture with respect to $H_0=\bla{\omega_0 +2\Omega_2\cos\Omega_1 t}\sigma_z/2$, with the transformation unitary 
\bml
U_0(t)=\exp \bla{-i \int_0^t H_0(t') dt'} \\
= \exp \bla{-i\blb{\frac{\omega_0}{2}+ \Omega_2\mbox{sinc}\Omega_1 t} \sigma_z  t}
\label{first_interaction_picture}
\end{multline}
we obtain (fig. \ref{time_dependent_concatenation})
\beq
H_I=-\Omega_2\sigma_z\cos\Omega_1 t +\delta B(t) \sigma_z +  \frac{\Omega_1}{2}\sigma_x+\frac{\delta \Omega_1(t) }{2}\sigma_x.
\label{1}
\eeq
which is achieved after using the RWA, $\Omega_1 \ll \omega_0$ and neglecting the counter-rotating terms. At this stage, we have reached the same point as the regular concatenated dynamical decoupling approach with two driving fields, although we only used a single one. Thus, the third term of Eq. \ref{1} is the energy gap, which is opened in the dressed state basis, and protects against the ambient magnetic field noise (the second term of Eq. \ref{1}). The first term of Eq. \ref{1} behaves like the second driving field that protects against the Rabi frequency noise (the last term of Eq. \ref{1}), as can be seen by following the derivation in the previous section. Unlike regular concatenated dynamical decoupling, the current protection (the first term of Eq. \ref{1}) originates from the time-dependent detuning, where the noise is assumed to be negligible. Hence there is no need to continue to concatenate further protecting driving fields. Note that a naive approach in which the AWG generates the time dependent phase and imprints it onto a resonant source $(\Omega_1 \cos \omega_0t)$, would introduce an enhanced amplitude noise $\delta \Omega_2(t)$. Rather, the AWG must generate the whole driving function; namely, the time dependently detuned driving field (last term of Eq. \ref{time-dependent}). In this way, the amplitude noise of the RWA gives rise to an enhanced $\delta\Omega_1(t)$ alone, which is compensated for. However, it does not affect $\delta\Omega_2(t)$, which as a result, is reduced.

%is not affected by  the does not interfere with the desired function, and thus, 
% {\bf What did you mean here?} ALEX

%\section{estimation of the time-dependent detuning resolution}
 
\section{Numerical simulation}
\begin{figure}[h]
\begin{centering}
\includegraphics[width=0.45\textwidth]{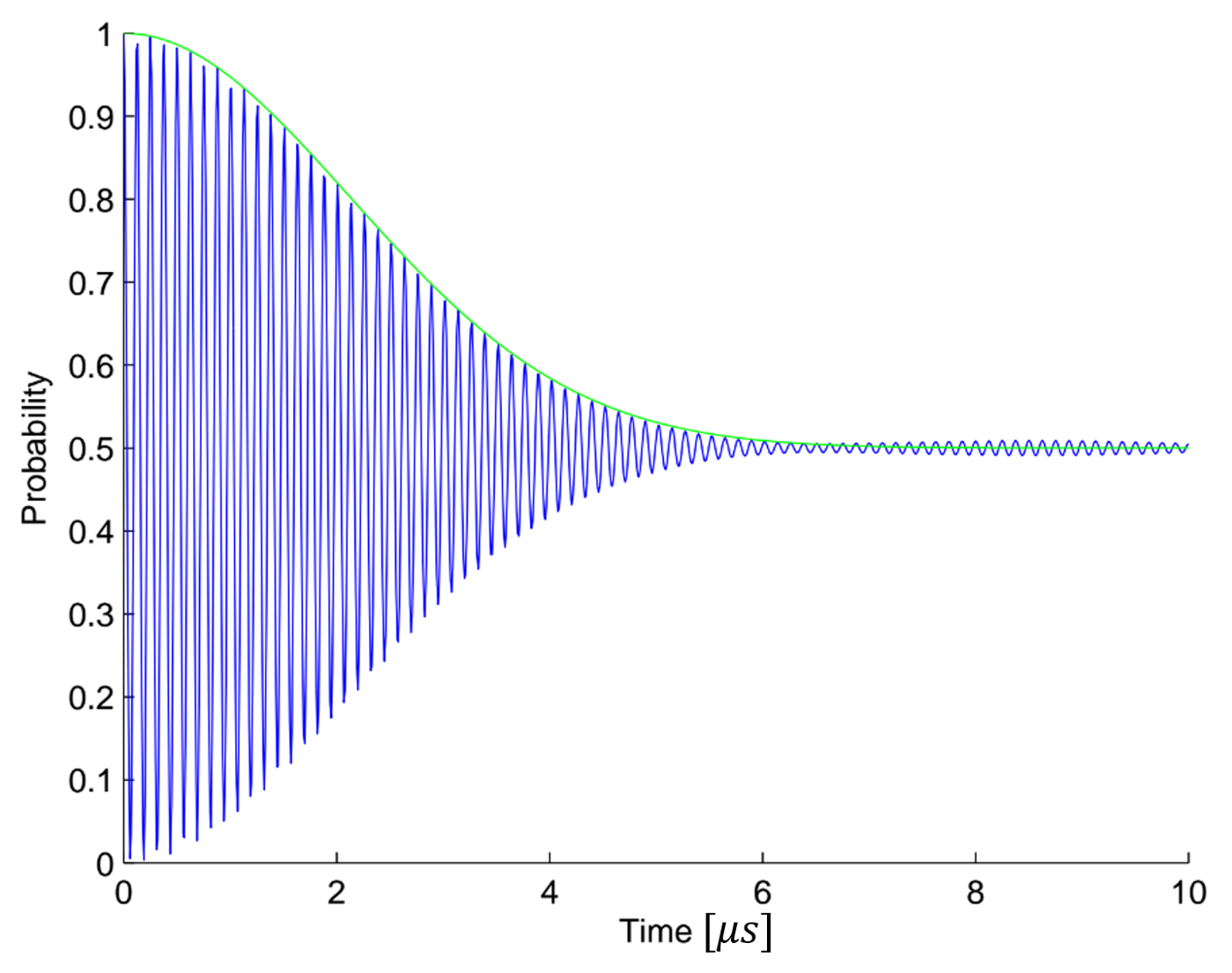}
\par\end{centering}
\protect\caption{Pure dephasing with no driving. The theoretical prediction  $\frac{1+e^{-\frac{g^{2}t^{2}}{2}}}{2}$
is plotted in green $\left(g^{2}=\frac{2}{T_{2}^{*}{}^{2}}\right)$. The excat OU process is plotted in blue after avareging over 1500 trails. Time in units of $\mu$s.}
\label{FigNS1}
\end{figure}

We numerically verified that our scheme results in an improved coherence time, by simulating its implementation with a two-level system
that is subject to both magnetic noise and power fluctuations of the driving field.
The noise, $B\left(t\right)$, was simulated as an Ornstein-Uhlenbeck (OU) process \cite{OU1,OU2} with a zero expectation value, $\left\langle B\left(t\right)\right\rangle =0$,
and a correlation function $\left\langle B\left(t\right)B\left(t^{'}\right)\right\rangle =\frac{c\tau}{2}e^{-\gamma\left|t-t^{'}\right|}$, 
where $c$ is the diffusion constant and $\tau = \frac{1}{\gamma}$ is the correlation time of the noise.
The OU process was realized by an exact algorithm \cite{OU3}, which according to
\begin{equation}
B(t+\Delta t)=B(t)e^{-\frac{\Delta t}{\tau}}+n\sqrt{\frac{c\tau}{2}\left(1-e^{-\frac{2\Delta t}{\tau}}\right)},
\end{equation}
where $n$ is a unit Gaussian random number. We used a pure dephasing
time of $T_{2}^{*}=3\:\mu s,$ and a correlation time of the noise of  $\tau=25\:\mu s$ (diffusion constant is 
given by $c\approx\frac{4}{T_{2}^{*}{}^{2}\tau}).$ 
An OU process was also used to realize driving fluctuations. Here we used a correlation time of
$\tau_{\Omega}=500\:\mu s$, and a relative amplitude error of $\delta_{\Omega}=0.5\%$  (diffusion constant is given by $c_{\Omega}=2\delta_{\Omega} \big/ \tau_{\Omega}$).
In case that dynamical decoupling is not employed (no driving fields),
the Hamiltonian is given by
\begin{equation}\label{H0d}
H=\frac{\omega_{0}}{2}\sigma_{z} + \frac{B(t)}{2}\sigma_{z}.
\end{equation} 
Our simulation agrees with the theoretical model and shows that this 
results in a pure dephasing time of $T_{2}^{*}=3 \mu$s (see Fig. \ref{FigNS1}).

We continued by adding a single on-resonance driving field with a Rabi frequency of $\Omega_{1}=10$ MHz. In this case we have that
\begin{equation}\label{H1d}
H=\frac{\omega_{0}}{2}\sigma_{z} + \Omega_{1}(1+\delta_{1}(t))\cos(\omega_{0}t)\sigma_{x} + \frac{B(t)}{2}\sigma_{z},
\end{equation}
were $\Omega_1 \delta_{1}(t)=\delta\Omega_1(t)$ in Eq. \ref{second_gap}. 

\begin{figure}[h]
\begin{centering}
\includegraphics[width=0.45\textwidth]{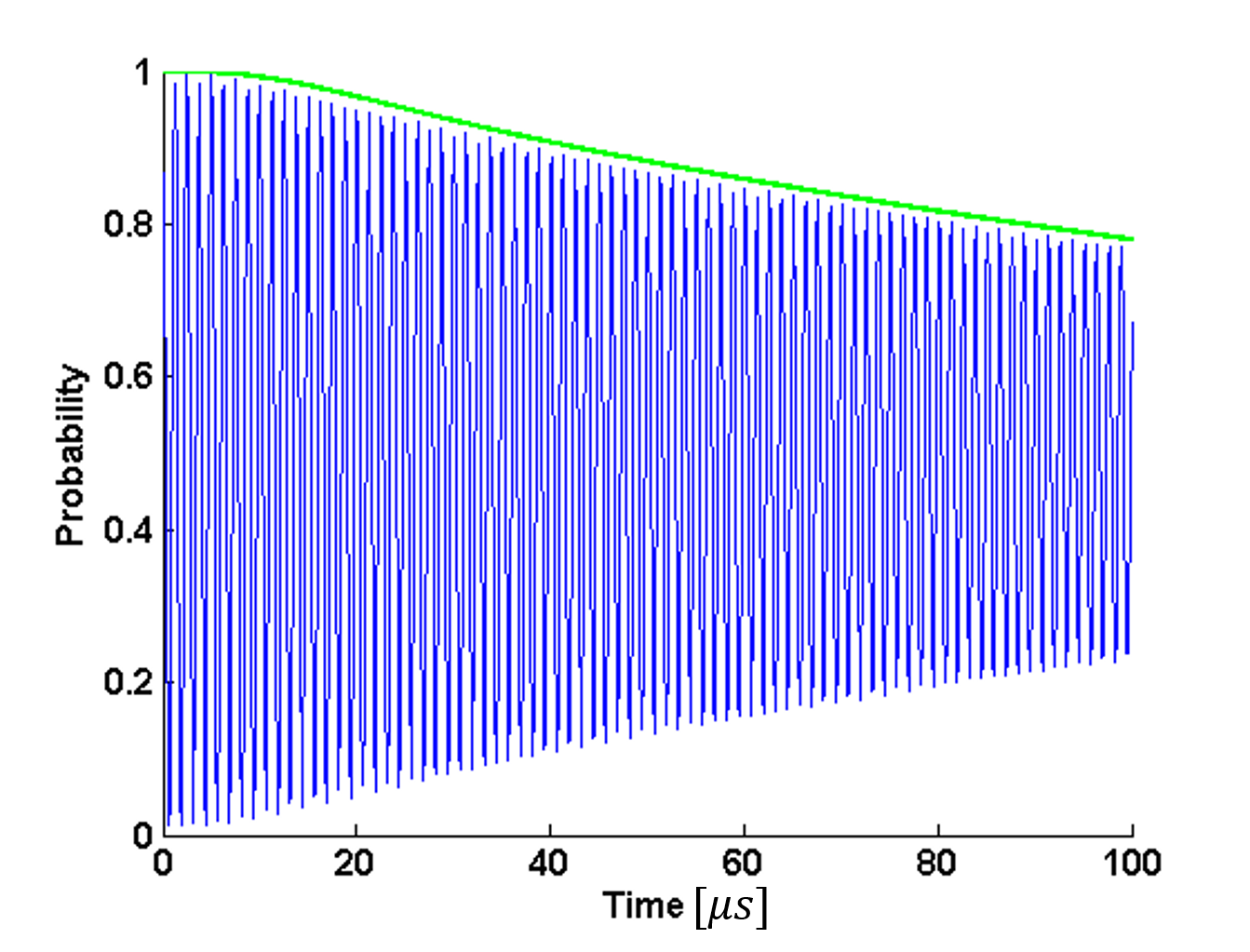}
\par \end{centering}
\protect\caption{Dephasing due to magnetic noise under an ideal single-drive. Analytical solution is plotted in green and corresponds to a coherence time of $T2=170 \mu$s. The excat OU process is plotted in blue after avareging over 2500 trails. Time in units of $\mu$s.}
\label{FigNS2}
\end{figure}

\begin{figure}[h]
\begin{centering}
\includegraphics[width=0.45\textwidth]{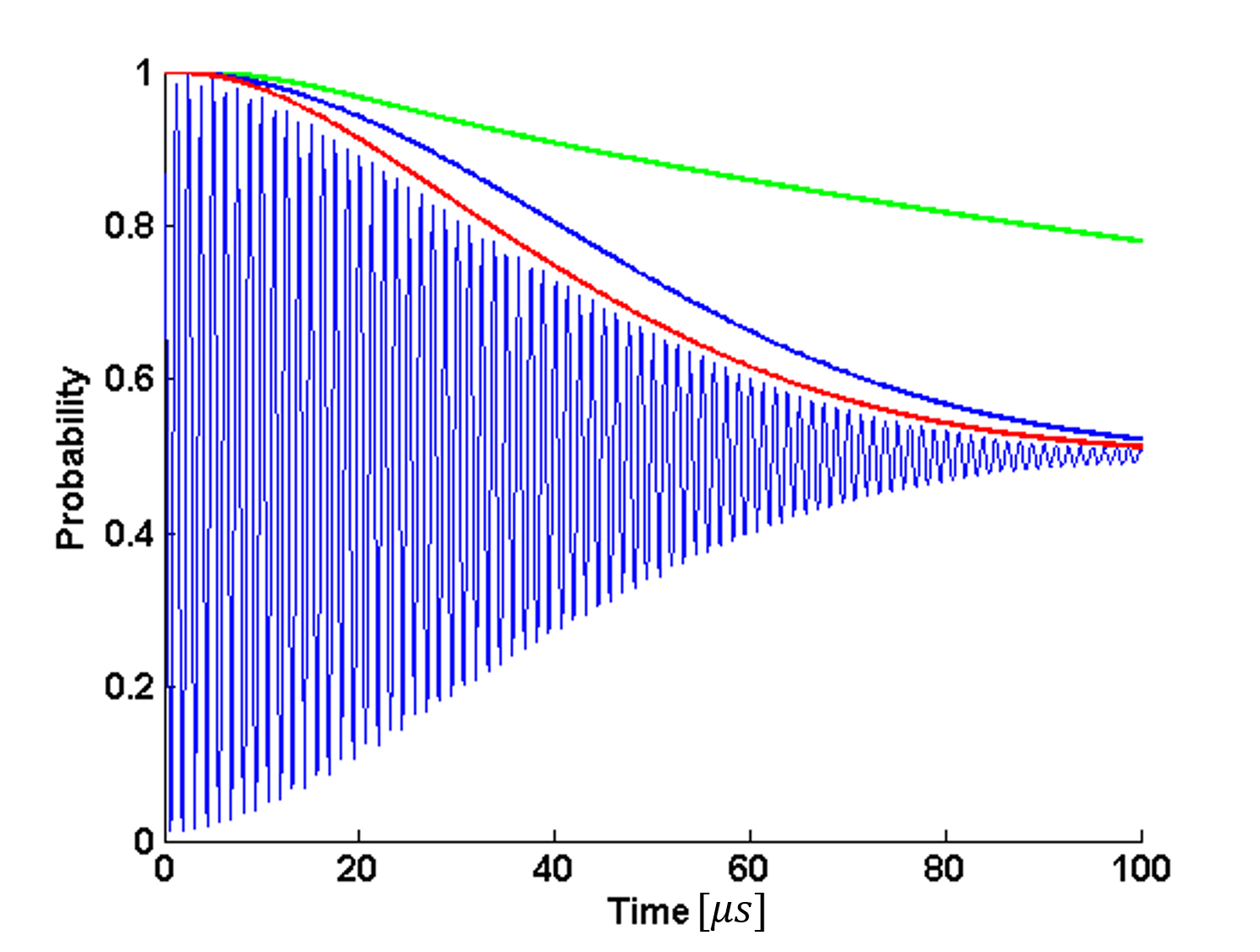}
\par \end{centering}
\protect\caption{Dephasing due to magnetic noise and driving fluctuations of a single-drive. Analytical solutions: (i) Dephasing due to magnetic noise (green), (ii) Dephasing due to driving noise (blue), and (iii) Total dephasing (red), which corresponds to a coherence time of $T2=57 \mu$s. The excat OU process is plotted in blue after avareging over 2500 trails. Time in units of $\mu$s.}
\label{FigNS3}
\end{figure}

In Fig. \ref{FigNS2}  (Fig. \ref{FigNS3}) the result of a simulation without (with) driving fluctuations is shown.
The analytical function of the dephasing rate due to magnetic noise under a strong driving field is known \cite{Aharon,Dobrovitski}, and in this
case the coherence time is given by $T2=170 \mu$s (green line in Fig. \ref{FigNS2}). Driving fluctuations, however, constitute the major source of dephasing. The coherence time due to driving fluctuations alone would be $T2=57 \mu$s (blue line in Fig. \ref{FigNS2}) and together with the effect of the magnetic noise, a single drive obtains $T2=50 \mu$s (red line in Fig. \ref{FigNS2}).

In case of a concatenated double-drive, driving fluctuations of the second driving field constitute the major source of dephasing. The Hamiltonian is given by

\begin{eqnarray}\label{H2d}
H &=& \frac{\omega_{0}}{2}\sigma_{z}  + \frac{B(t)}{2}\sigma_{z} \nonumber\\
 &+&\Omega_{1}(1+\delta_{1}(t))\cos(\omega_{0}t)\sigma_{x}\nonumber\\
 &+&\Omega_{2}(1+\delta_{2}(t))\cos(\frac{\Omega_{1}}{2}t)\sigma_{z},
\end{eqnarray}
were $\Omega_i \delta_{i}(t)=\delta\Omega_i(t)$ for $i=1,2$. 

\begin{figure}[h]
\begin{centering}
\includegraphics[width=0.45\textwidth]{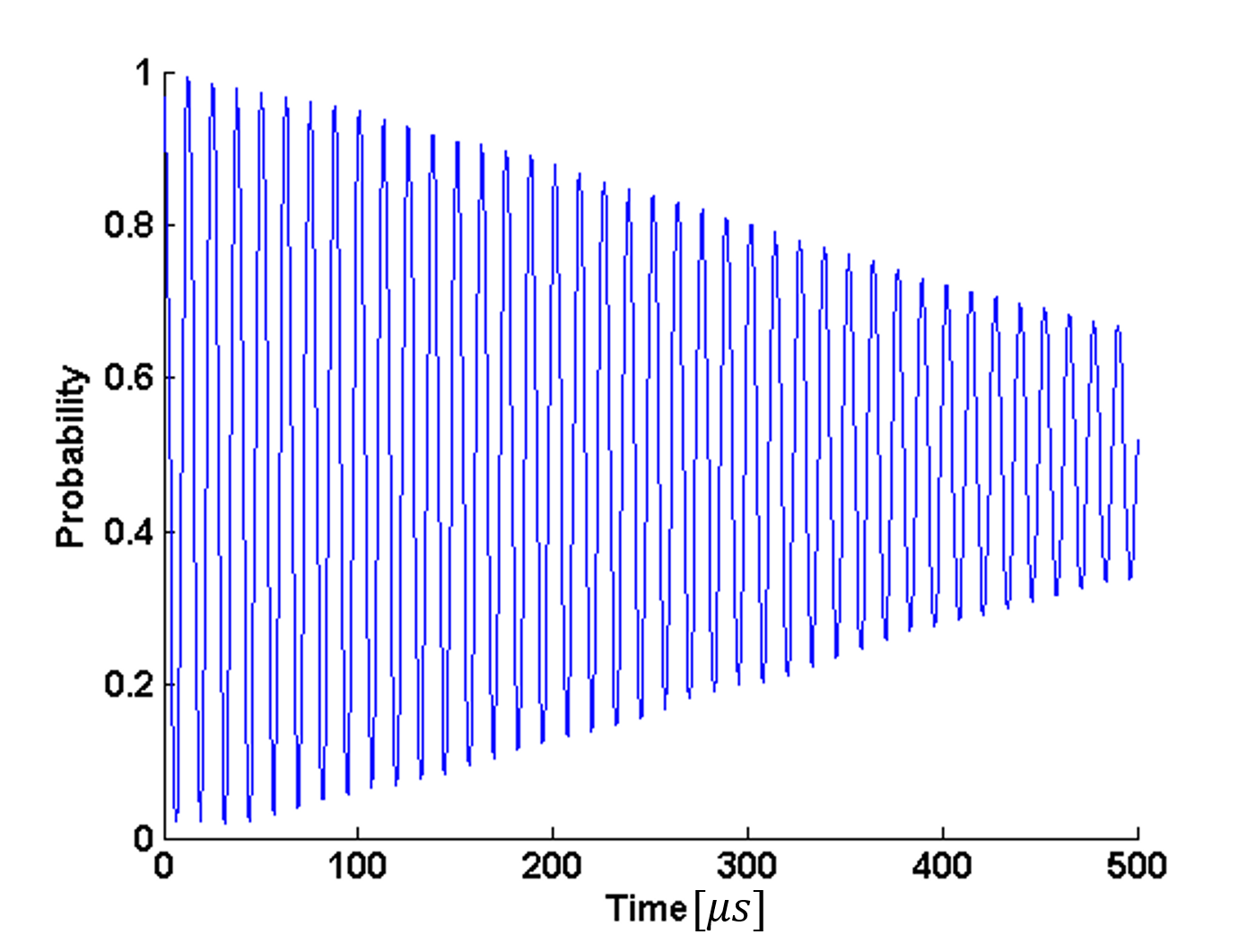}
\par \end{centering}
\protect\caption{Dephasing due to magnetic noise and driving fluctuations of a concatenated double-drive. Dephasing is mostly due to noise in the second driving field $\delta \Omega_2(t)$ which is not compensated for. The double-drive results in a coherence time of $T2\simeq 450 \mu$s. The excat OU process is plotted in blue after avareging over 1000 trails. Time in units of $\mu$s.}
\label{FigNS4}
\end{figure}

The simulation shows that the a concatenated double-drive results in $T2\simeq 450 \mu$s, where we used a second drive with
a Rabi frequency of $\Omega_{2}=1$ MHz (see Fig. \ref{FigNS4}).  

\begin{figure}[h!]
\begin{centering}
\includegraphics[width=0.45\textwidth]{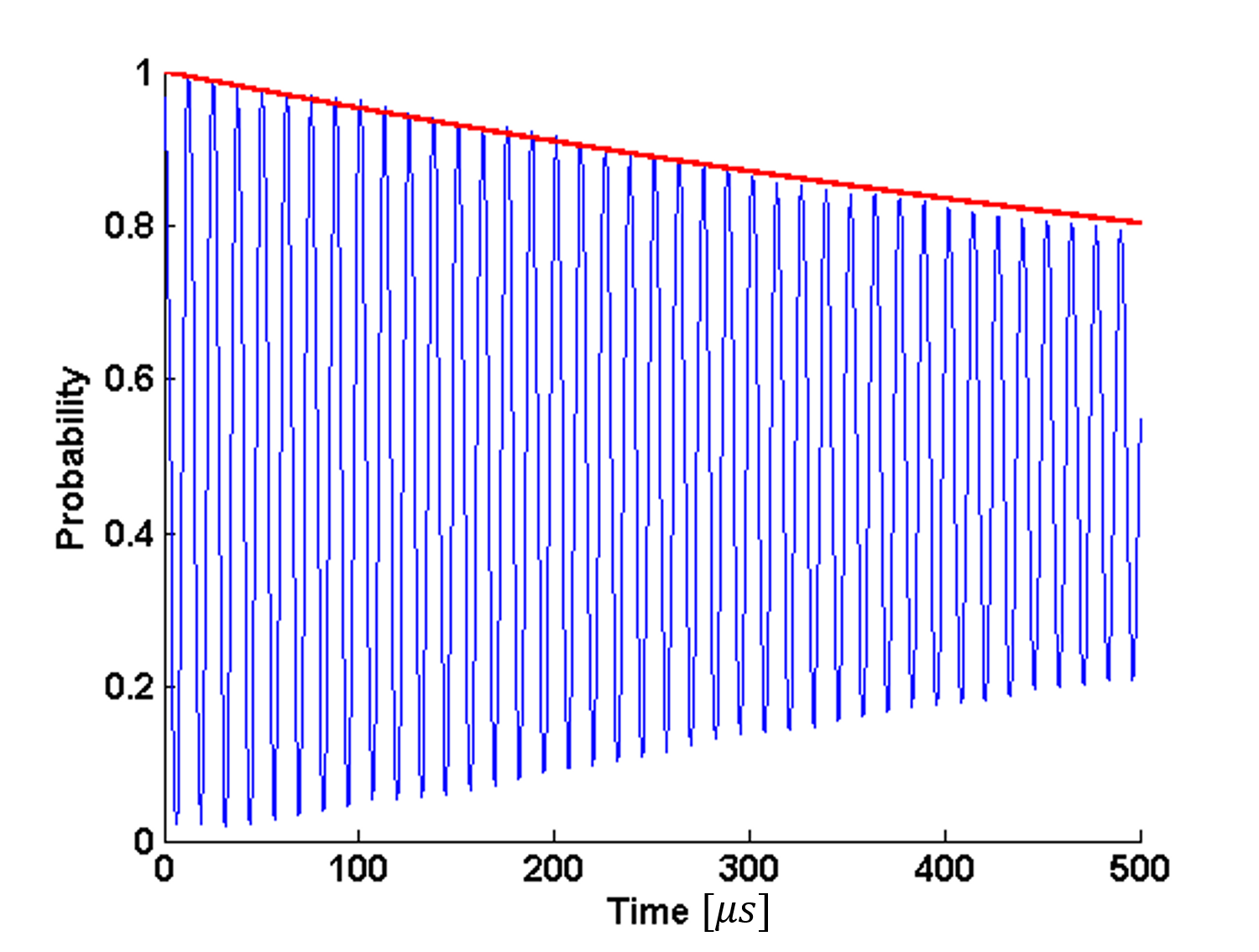}
\par \end{centering}
\protect\caption{Dephasing due to magnetic noise and driving fluctuations of the time-dependently detuned driving field. Assuming $\delta \Omega_2(t)=0$, dephasing is mostly due to noise $\delta \Omega_1(t)$ which is compensated for. This approach results in a coherence time of $T2\simeq 1000 \mu$s. $\frac{1+e^{-t/1000}}{2}$ is plotted in red. The excat OU process is plotted in blue after avareging over 1000 trails. Time in units of $\mu$s.}
\label{FigNS5}
\end{figure}

In our scheme there is no first order dephasing rate of the doubly-dressed states because there are no fluctuations in $\Omega_{2}$. 
The dephasing now comes mostly as a second order effect of the fluctuations in $\Omega_{1}$. Indeed, our simulation suggests that in this case a coherence time of  $T2\simeq 1000 \mu$s is attained (see Fig. \ref{FigNS5}). \\

\section{Utilizing the time dependent detuning for Magnetometry}
In a magnetometery experiment, a signal whose amplitude is related to the magnetic field is measured. In the following, we propose two ways to perform magnetometry experiments while utilizing the protecting time dependently detuned driving field to increase the coherence time, which results in a better sensitivity.

\subsection{First magnetometry approach}
The first magnetometry approach is realized by measuring a magnetic field polarized in the $z$ direction with frequency $\omega_d$. Thus, the Hamiltonian of the bare energy, the protecting time dependently detuned driving field (Eq. \ref{time-dependent}), and the $z$ signal reads %{\bf This sentence is not clear} ALEX
\beq
H=\frac{\omega_0}{2} \sigma_z + \Omega_1\sigma_x \cos \bla{\blb{\omega_0 +2\Omega_2\mbox{sinc}\Omega_1 t}t} + g \sigma_z \cos \omega_d t, 
\eeq
where we have omitted the noisy terms.

Following the same derivation steps, by first moving to the first interaction picture (Eq. \ref{first_interaction_picture}), the signal is not affected, and we obtain
\beq
H_I=-\Omega_2\sigma_z\cos\Omega_1 t +  \frac{\Omega_1}{2}\sigma_x +  g \sigma_z \cos \omega_d t.
\eeq

In the second stage we move to the interaction picture with respect to the dressed state energy $ \frac{\Omega_1}{2}\sigma_x$, which protects against the magnetic noise. Thus, after using the RWA, assuming $\Omega_1 - \omega_d\ll\Omega_1 + \omega_d$, we obtain
\beq
H_{I_2}=-\frac{\Omega_2}{2}\sigma_z  +  \frac{g}{4} \blb{\bla{\sigma_z -i\sigma_y} e^{i (\Omega_1 - \omega_d)t}  + h.c}.
\eeq
In this stage we have two optional ways to proceed. If the resonant condition holds $\Omega_1 = \omega_d$, we can use a Ramsey experiment to measure a shift $g/2$ in the double dressed state energy. If we set the detuning $\Omega_1 - \omega_d=\pm \Omega_2$, we can perform the Rabi experiment in the double dressed states, with Rabi oscillations $g/4$. Both alternatives use the RWA where $g \ll \Omega_2$.

\subsection{Second magnetometry approach}
The second approach to realizing a magnetometry experiment while utilizing the time dependently detuned driving field uses an oscillating magnetic field, polarized in the $x$ direction. Unlike the first approach, now the magnetic fields that we sense should be oscillating fast, to survive the bare energy gap. %Note that sensing high frequency signals is challenging when using pulsed dynamical decoupling schemes, because the time separation between the dynamical decoupling pulses  has to be short $\Delta t = 2\pi/\omega_d$. However, the continuous dynamical decoupling approach overcomes this challenge. 
The Hamiltonian containing the bare energy, the protecting time dependently detuned driving field (Eq. \ref{time-dependent}), and the $x$ signal reads
\beq
H=\frac{\omega_0}{2} \sigma_z + \Omega_1\sigma_x \cos \bla{\blb{\omega_0 +2\Omega_2\mbox{sinc}\Omega_1 t}t} + g \sigma_x \cos \omega_d t, 
\eeq 
neglecting the noisy terms.

Following the same derivation steps as above, we first move to the first interaction picture (Eq. \ref{first_interaction_picture}), where the signal is affected, and we obtain
\beq
H_I=-\Omega_2\sigma_z\cos\Omega_1 t +  \frac{\Omega_1}{2}\sigma_x +  \frac{g}{2} \bla{\sigma_+f(t) +h.c}.
\label{magnet_x}
\eeq
where we have used the RWA assuming $\omega_0 -\omega_d\ll\omega_0 +\omega_d$, and the exponent function is
\beq
f(t) = e^{i (\omega_0- \omega_d +2\Omega_2\mbox{sinc}\Omega_1 t) t}.
\eeq

In the next stage of the derivation we move to the interaction picture with respect to the dressed state energy $ \frac{\Omega_1}{2}\sigma_x$, which protects against the magnetic noise. Here it is easier to rotate the system by $-\pi/4$ around the $y$ axis, such that $\sigma_x \rightarrow S_z$, $\sigma_z \rightarrow -S_x$, and $\sigma_y \rightarrow S_y$. Thus we obtain
\beq
H_{I_2}  = \frac{\Omega_2}{2}S_x + \frac{g}{4} \blb{\bla{S_z + S_+ e^{i\Omega_1 t} - S_- e^{-i\Omega_1 t}} f(t) + h.c}.
\eeq

Next, we move to the frame rotating with the double dressed state energy gap $\Omega_2 S_x/2$. For convenience, we rotate the system back with $\pi/4$ around the $y$ axis; namely, we transform back to the $\sigma$ basis. Therefore, we obtain
\bea H_{I_3} &=& \frac{g}{4} \bla{\sigma_+ e^{i\Omega_2 t} +\sigma_-e^{-i\Omega_2 t} }  f(t) +h.c \\
&+& \frac{g}{8}\blb{-\sigma_z +\sigma_+e^{i\Omega_2 t} -\sigma_- e^{-i\Omega_2 t} }  e^{i \Omega_1 t} f(t)  + h.c \\
&-& \frac{g}{8}\blb{-\sigma_z -\sigma_+e^{i\Omega_2 t}+\sigma_- e^{-i\Omega_2 t} }  e^{-i \Omega_1 t} f(t)  + h.c 
\label{third frame}
\eea

Since $\Omega_2 \ll \Omega_1$, we can approximate the exponent function
\beq 
f(t) \approx e^{i (\omega_0- \omega_d)t}\blb{1+ i\frac{2\Omega_2}{\Omega_1}\sin\Omega_1 t }.
\eeq

Similar to the first magnetometry approach, we also have two ways to proceed. By setting the detuning $\omega_0 - \omega_d=\pm\Omega_2$, we are left with the following energy preserving terms
\beq
H_{I_3} =   \frac{g}{4} \sigma_x +O\bla{\blb{\frac{\Omega_2}{\Omega_1}}^2}, 
\eeq
which can be used for a Rabi experiment in the double dressed state basis. The second way to proceed from Eq. \ref{third frame} is to set the detuning $\omega_0 - \omega_d=\pm\Omega_1$, thus resulting in 
\beq
H_{I_3} =   \pm\frac{g}{4} \sigma_z +O\bla{\blb{\frac{\Omega_2}{\Omega_1}}^2}, 
\eeq
which can be used for a Ramsey experiment in the double dressed state basis.

%{\bf In the second method it should be stresses that high frequency fields could be read in contrast to regular decoupling. This is a very important advantage.} ALEX

%\section{Utilizing time dependent detuning for entangling gates in trapped ions and NVs}

%To generate entanglement between spins one should induce coupling between  

%Coupling spins in order to entangle them is usually time consuming. During the relatively long duration of the entanglement process the system is vulnerable to noise and decoherence. For this reason, dynamical decoupling techniques are used to reduce the influence of noise on quantum entanglement. Here, we propose using the time-dependently detuned driving field as an enhanced protection against noise and decoherence during the entanglement process. In particular we explain how our dynamical decoupling method can be implemented in systems of trapped ions and nitrogen-vacancy (NV) in diamond.      

\section{conclusion and summary}
In this manuscript we propose a new scheme for continuous dynamical decoupling, which removes the complexity of the regular scheme, and possesses a reduced Rabi frequency noise. By imposing a time-dependent detuning, our single driving field opens energy gaps in both the dressed state and the double dressed state basis. Thanks to the enhanced accuracy of the AWG, the fluctuations of the double dressed state energy gap are further reduced compared to the regular continuous scheme. We show how the new technique can be used for magnetmetry, where the enhanced coherence time gives rise to a better sensitivity. Furthermore, we note that this new dynamical decoupling scheme might be utilized for increasing the coherence time and the fidelity of dressed state based entangling gates \cite{Lemmer2013njp,Itsik2015NJP} and dressed state based quantum simulations \cite{Itsik_Haldane1,Itsik_Haldane2,Cai2015prl}, for both trapped ions and nitrogen-vacancy defects in diamond.

\section{acknowledgments}

We acknowledge the support of the Israel Science Foundation (grant no. 039-8823), the European commission (STReP EQUAM Grant Agreement No. 323714), EU Project DIADEMS, the Marie Curie Career Integration 6 Grant (CIG) IonQuanSense(321798) and the Niedersachsen-Israeli Research Cooperation Program. This work was partially supported by the US Army Research Office under Contract W911NF-15-1-0250.

%%%%%%%%%%%%%%%%%%%%%%%%%%%%%%%%%%%%%%%%%%%%%%%%%%%%%%%%%%%%%%%%%%%%%%%%%%%%%

\end{document}